\documentclass[12pt]{article}

\usepackage{times} 
\usepackage[T1]{fontenc}
\usepackage[utf8]{inputenc}
\usepackage[margin=1in]{geometry} 
\usepackage{setspace}
\usepackage[style=apa]{biblatex}
\usepackage[hidelinks]{hyperref}
\usepackage{graphicx}
\usepackage{pgfplots}
\usepackage{subcaption} % <-- Add this package for subfigures
\pgfplotsset{compat=1.18} % Use a recent compatibility version
\usepackage[table]{xcolor} % For coloring cells, rows, and columns
\usepackage{makecell}      % For multi-line cells in the header
\usepackage{placeins}
\usepackage[absolute,overlay]{textpos}

\begin{document}

% --- TITLE PAGE / COVERING PAGE ---
\begin{titlepage}
    \centering
    \vspace*{\fill}
    
    {\huge\bfseries Shifting landscape of disability and development in India: Analysis from historical trends to future
    predictions 2001-2031\par}
    
    \vspace{2.5cm} % Increased spacing for better balance
    
    % Use minipages for side-by-side author information
    \begin{minipage}{0.45\textwidth}
        \centering
        {\large\bfseries Hana Kapadia\par}
        \vspace{0.3cm}
        {\large Research Student at Cambridge Centre for International Research \par}
        \textit{Email: kapadiahana@gmail.com\\
        Contact Number: (+1) 650-924-7444}
    \end{minipage}
    \hfill % This command creates space between the two minipages
    \begin{minipage}{0.45\textwidth}
        \centering
        {\large\bfseries Arun Kumar Rajasekaran*\par}
        \vspace{0.3cm}
        {\large Research Scholar, University of Cambridge\par}
        \textit{Email: rarunrd@gmail.com*\\
        Contact Number: (+44) 758-626-4260}
    \end{minipage}
    
    \vspace{2.5cm} % Increased spacing for better balance
    
    {\large\bfseries \par}
    \vspace{0.5cm}
    \begin{minipage}{0.8\textwidth}
        \centering
        \itshape
        TISS Journal of Disability Studies and Research (TJDSR) Volume-V, Issue-II, December 2025

        \url{https://tiss.ac.in/uploads/files/TISS_Journal_FINAL-1_compressed_HHOqB3N.pdf}
    \end{minipage}
    
    \vspace*{\fill}
\end{titlepage}

\begin{abstract}
This study delves into the causes and trends of disability-related health burdens across Indian states. Through multiple Disability-Adjusted Life Years (DALY) types (covering communicable diseases, noncommunicable diseases, and injuries), gender disparities, and Human Development Index (HDI) values, these disability trends were evaluated. The data for this study was compiled from censuses, health research organisations, and data centres, among various other sources. We built regression  models and used them to analyze trends across past decades and make projections for 2031. Our regression results show a strong inverse relationship between communicable disease DALYs and HDI. In other words, ongoing improvements in development and infrastructure significantly reduced communicable disease DALYs. In contrast, noncommunicable DALYs did not decrease despite rising HDI. And lastly, injury DALYs showed moderate declines with higher HDI, which reflects improvements in healthcare and safety systems. Gender analysis showed male overrepresentation among people with disabilities. These results from our study support that there is a need to shift public health focus toward chronic diseases and address gender disparities in disability outcomes.

 \vspace{0.5cm} % Increased spacing for better balance

\\
\\
\\
\\
\\
\\
\\
\end{abstract}

% --- Corresponding Author Block ---
% This is the snippet you requested. It must be placed after \maketitle
% within the document environment.
% You may need to adjust the vertical position (the `4.5in` value) to
% ensure it sits correctly after your keywords on the first page.
\begin{textblock*}{\textwidth}(0.5in, 6in) % {width}(x-position, y-position)
    \noindent \textbf{Corresponding Author:}\\
    Arun Kumar Rajasekaran\\
    Research Scholar, University of Cambridge,\\
    Hughes Hall, Wollaston Road, \\
    Cambridge CB1 2EW, United Kingdom \\   
    Phone: [(+44) 758-626-4260]\\
    Email: rarunrd@gmail.com
\end{textblock*}

\section{Introduction}
India has a population of over 1.4 billion people. With this large population and regional inequalities, understanding disability burdens and trends is especially necessary (Rashmi \& Mohanty, 2024).  Disability plays a large role in the overall health of communities, especially in social and economic aspects. Disability, as defined by The World Health Organization, is the relationship between health conditions and outside environmental factors. The impact of disability is best captured through Disability-Adjusted Life Years (DALYs). DALYs are a measure of disease burden and combine premature mortality and years lived with disability into a numerical value (Murray \& Acharya, 1997). In recent decades, India has experienced increased development. This can be easily interpreted from the rising Human Development Index (HDI) values, improved education, and greater access to healthcare (Smits \& Permanyer, 2019; United Nations Development Programme, 2025). Yet, while communicable diseases have decreased significantly, noncommunicable diseases and injury-related burdens remained challenges (Murray \& Acharya, 1997). Gender also affects disability reporting. The societal risks and biases that both men and women face can affect data and policy responses (Registrar General and Census Commissioner, India, 2014a, 2014b).\\

This paper aims to address the following objectives \\
1. Analyze how DALY burdens have evolved across Indian states in the past decades (2001-2011) and how they relate to HDI\\
2. Determine the gender disparities in disability representation and what factors contribute to them\\
3. Use historical data to develop a machine learning model that identifies and predicts DALY and HDI trends in present and future (2021-2031)\\

By addressing these targets, through this study we aim to provide insights into the changing landscape of disability in India, highlight areas of persistent inequality, and inform policies that address both emerging and present health challenges.

\section{Literature Review}

Types of publications on health and disability trends in India that provide context for this study are those that document India’s epidemiological transition, analyze the connections between socioeconomic development and health outcomes, or examine the effect of gender on disability.

Prior to 2016, communicable diseases accounted for the majority of India's total disease burden. However, that year, noncommunicable diseases accounted for 61\% of India's total disease burden, though this wasn’t true across all states, with variance ranging from 75\% in Kerala to just 48\% in Bihar (India State-Level Disease Burden Initiative Collaborators, 2017). The DALY metric was critical in this analysis (Murray \& Acharya, 1997). It is pertinent to note that, as measured by DALYs, the health loss from diabetes was more rapid than any other noncommunicable disease (from 26 million in 1990 to 65 million in 2016.) This rapid growth was linked to high BMI (overweight), and displays how diabetes is a major public health crisis (India State-Level Disease Burden Initiative Diabetes Collaborators, 2018). Issues such as suicide deaths are also a major component of the injury burdens (Dandona et al., 2020).

Socioeconomic development and health are often heavily related. The study from Rashmi and Mohanty (2024) implores that locomotor, mental, and multiple disabilities are higher among the poorest households compared to the wealthiest. This displays the relationship between poverty and disability. The poorest states and social groups often bear a higher burden of disability. Low socioeconomic status is also an accurate predictor of cardiovascular diseases, comparable to other common risk factors such as hypertension, diabetes, and smoking (Schultz et al., 2018). Lower-caste (Dalit) disabled women also experience a "triple form of discrimination" based on their gender, caste, and disability. The systematic oppression they face often limits their access to healthcare and leaves them overlooked in academic literature, causing them to be invisible and vulnerable (Maurya 2023). This is due to lower income being linked with less education, lower quality of care, less access to healthy food, and fewer spaces for physical activity.

\section{Methodology}

\subsection{Data collection for past and current years}
The data used in this study was compiled from the 2001 and 2011 India Census conducted by the Government of India (Registrar General and Census Commissioner, India, 2014a; 2014b), the Institute for Health Metrics and Evaluation (IHME) (Tichenor, \& Sridhar, 2020), and Global Data Lab, an independent data and research center at Nijmegen School of Management of Radboud University (Smits \& Permanyer, 2019). From the census data, we used the number of disabled individuals per state, total and separated by gender. From IHME we compiled Disability-Adjusted Life Years (DALY) values, and from Global Data Lab, we utilized the Human Development Index (HDI) values.

This data was chosen because they provide a complete picture of how human development and other socioeconomic factors affect disease burden. By utilizing multiple sources and seperating data by Indian state, the compiled dataset was able to determine the determining factors for wellbeing in a society. The gender data displays how men and women experience different rates of disability and healthcare access (Registrar General and Census Commissioner, India, 2014a; 2014b). The final dataset had 29 areas (28 states and 1 total India measurement), 6 features, and four decades. Union territories Andaman \& Nicobar Island, Dadra and Nagar Haveli, Daman \& Diu, Lakshadweep, Puducherry, and Chandigarh were not included in this study due to unavailability of DALY data. Odisha (formerly Orissa) was also excluded due to its lack of 2001 census data.\\

The six features, compiled to study each Indian state are:\\
- DALY index, Type-A, (values from 2001, 2011, 2021, projected 2031)\\
- DALY index, Type-B, (values from 2001, 2011, 2021, projected 2031)\\
- DALY index, Type-C, (values from 2001, 2011, 2021, projected 2031)\\
- HDI (values from 2001, 2011, 2021, projected 2031)\\
- Disabled M/F Ratio (values from 2001, 2011, projected 2021, projected 2031)\\
- Total M/F Ratio (values from 2001, 2011, projected 2021, projected 2031)\\

DALY is a measure of overall disease burden, determined through the sum of years lost of life due to premature mortality and years lived with the disability. It quantifies the burden that disabilities cause in daily life. DALYs will decrease if the impact of disabilities lessens and will increase with the contrary. In the data, to ensure consistency throughout different population sizes, DALYs were represented as a rate per 100,000 people in the population. \\

In this study, DALY was broken down into three key components: \\
1. Type-A: Communicable, maternal, neonatal, and nutritional diseases \\
2. Type-B: Noncommunicable diseases (Neoplasms, Mental, Respiratory, etc.)\\
3. Type-C: Injuries (Unintentional, Accident, Self-harm, etc.)\\

Human Development Index (HDI) values measure overall national/subnational achievements. The calculation is made using three main dimensions: Life expectancy, education (years of school), and standard of living (Gross National Income per capita). It is measured between 0.00 and 1.00, with a low value being under 0.550, medium 0.0550-0.699, high 0.700-0.799, and very high above 0.800.

\subsection{Data predictions for future analysis}

The past data were used to predict DALY and HDI values for 2031. This was done in four sequential steps: Feature Construction, Model Calibration, HDI prediction, and finally DALY Prediction. 

A dataset suitable for machine learning was first assembled. The objective was to create a consolidated dataset for which the regression model could utilize. The dataset included explanatory variables, the features.

The model used different input variables depending on the prediction:\\
- Predicting 2031 HDI: HDI (values from 2001, 2011, and 2021)\\
- Predicting 2031 DALY Communicable: HDI (values from 2001, 2011, 2021, and predicted 2031), and DALY  Communicable (values from 2001, 2011, and 2021)\\
- Predicting 2031 DALY Noncommunicable: HDI (values from 2001, 2011, 2021, and predicted 2031), and DALY Noncommunicable (values from 2001, 2011, and 2021)\\
- Predicting 2031 DALY Injury: HDI ( values from 2001, 2011, 2021, and predicted 2031), and DALY Injury (values from 2001, 2011, and 2021)\\

Treating the three DALY metrics as separate features instead of summing them allowed the model to differentiate the specific influences of each type of health challenge on overall human development. For instance, it can quantify whether improvements in sanitation (affecting communicable DALYs) have a greater or lesser impact on HDI than improvements in road safety (injury DALYs). It also provides more variables for the model to analyze, allowing more data to back its predictions. The following subsections go into further details of ML techniques employed to attain the future projected values used in this study. 

\subsubsection{HDI  values prediction methodology}
Method: A Linear Regression based on Time Intervals was used. A linear regression was performed on the data points from the years 2001, 2011, and 2021 to find the best fit model that represents the average rate of HDI growth over this period. This model was then extended to project the HDI value for 2031. 

This projection of 2031 Human Development Index (HDI) is used and is the foundation of all subsequent Disability-Adjusted Life Years (DALY) forecasts. Because of this, a linear regression model was chosen for its reliability. This model was chosen over more complex alternatives such as polynomial and logarithmic because of the risk of overfitting and instability, especially with only three data points. To avoid error propagation into further predictions, this stable and defensible baseline was utilized. 

Note: These predictions are based on linear extrapolation-based modelling and hence do not account for non-linear trends, such as decreased rates of HDI growth or the impacts of unforeseen economic/social events. Nevertheless, this simple model seems to capture enough trends, and help us in exploring the dataset in question. 

\subsubsection{DALY values prediction methodology}
For each of the three DALY categories—Communicable, Noncommunicable, and Injury—three models (Linear, Polynomial (Quadratic) Regression, and Exponential Decay) were tested to determine the best fit for the data. An R squared test was performed to measure the fit.

Modelling DALY - Type A (Communicable)  
The Exponential Decay model was chosen because it had the second-strongest statistical fit to the data with nearly perfect R-squared values across most states, while also following the expected real world trend: as HDI continues to rise, further reductions in DALYs become increasingly difficult, leading to diminishing returns. The polynomial model, though having perfect R2 of 1.0, was discarded due to instability warnings (‘Polyfit may be poorly conditioned’), indicating a high risk of overfitting.

Modelling DALY - Type B (Noncommunicable) 
The Linear Regression model was chosen for a multitude of reasons. Firstly, the relationship between development and noncommunicable diseases is governed by the ‘epidemiological transition’, a theory that describes how as HDI increases and society develops, populations live longer, leading to an increase in chronic conditions and noncommunicable diseases. Because of this, the exponential model was deemed too risky and unstable, introducing too many unknowns and too much randomness. 
Again, the R-squared (R²) value was used to measure the goodness of fit and the polynomial model was discarded due to risk of overfitting. 
Due to these reasons, the Linear Regression model was selected. 

Modelling DALY - Type C (Injury)
A Linear Regression model was chosen. The R-squared analyses showed that both the Linear and Exponential models provided a strong fit to the data, while the Polynomial model was again discarded due to the risk of overfitting. Linear Regression was ultimately chosen because, although the Exponential also fit well, the linear approach is a more conservative choice. The exponential model assumes a pattern of diminishing returns, where injury reduction becomes harder as HDI increases, but there is no such theory that allows for this belief. The linear model removes this layer of complexity while still captures the downward trend without making strong and potentially incorrect assumptions about the shape of the relationship.

\section{Analytical Discussion}
\subsection{Analysis from the past decades (2001-2011)}
\subsubsection{DALY vs HDI Comparison Results}
The DALY and HDI measures are both numerical representations of a population’s overall well-being. The relationship between these two measures can be analyzed, allowing us to identify how economic developments and state systems can influence and shape disability burdens, then informing future policies. For example, improvements in infrastructure such as public transport can reduce disease transmission (lowering DALY), while also encouraging economic growth by increasing productivity. Understanding the interplay between DALY and HDI allows governments and allied organizations to design more effective interventions and understand all possible implications of their policies.\\

\textit{DALY type A - Communicable, maternal, neonatal, and nutritional diseases }\\
Through Figure 1, we can see how in both 2001 and 2011, even though there is considerable variability, states with higher HDI have consistently lower DALY values. As HDI increases, DALY values tend to decrease. However, the scatter in the 2001 data suggests that there are other factors also playing a role in determining the DALY values in the states. In 2011, a much clearer and stronger negative correlation between HDI and DALY was observed. States with lower HDI, around 0.50–0.55, still have higher DALY values, but the numbers are lower than in 2001 (24,000 vs. 38,000). By 2011, the plot shows a more pronounced downward trend, with states with higher HDI having lower DALYs than in 2001.

\begin{figure}[h!]
\centering % Center the entire figure

% --- First Subfigure (Left Plot) ---
\begin{subfigure}{0.49\textwidth}
    \centering
    \begin{tikzpicture}
    \begin{axis}[
        title={DALY vs HDI 2001 (Type A)},
        xlabel={HDI},
        ylabel={DALY per 100,000},
        grid=major,
        ymin=0,      % Sets the minimum y-axis value
        ymax=40000,   % Sets the maximum y-axis value
        scatter/classes={
            a={mark=o, fill=blue, draw=blue} % Define a class for blue circles
        }
    ]
    \addplot[
        scatter,
        only marks,
        scatter src=explicit symbolic % Use the symbolic class 'a' for all points
    ] table [x=HDI, y=DALY, meta=class] {
        DALY      HDI     class
        24986.39  0.495   a
        25749.53  0.478   a
        18780.06  0.499   a
        28111.71  0.486   a
        31633.82  0.432   a
        32879.71  0.554   a
        8703.15   0.615   a
        21741.16  0.526   a
        19931.57  0.546   a
        13257.39  0.589   a
        13669.25  0.53    a
        31677.04  0.554   a
        17480.04  0.516   a
        6449.2    0.604   a
        36002.52  0.456   a
        17442.08  0.556   a
        14535.66  0.556   a
        21661.84  0.447   a
        14004.55  0.571   a
        14091.6   0.519   a
        15300.31  0.658   a
        12871.92  0.575   a
        28739.27  0.466   a
        13451.64  0.546   a
        15095.22  0.618   a
        15336.77  0.526   a
        37617.04  0.46    a
        20681.1   0.62    a
        16716.87  0.502   a
    };
    \end{axis}
    \end{tikzpicture}
    \caption{Plot for the year 2001}
    \label{fig:daly2001}
\end{subfigure}
\hfill % This creates a flexible space between the two subfigures
% --- Second Subfigure (Right Plot) ---
\begin{subfigure}{0.49\textwidth}
    \centering
    \begin{tikzpicture}
    \begin{axis}[
        title={DALY vs HDI 2011 (Type A)},
        xlabel={HDI},
        ylabel={DALY per 100,000},
        grid=major,
        ymin=0,      % Sets the minimum y-axis value
        ymax=40000,   % Sets the maximum y-axis value
        scatter/classes={
            a={mark=o, fill=blue, draw=blue} % Define a class for blue circles
        }
    ]
    \addplot[
        scatter,
        only marks,
        scatter src=explicit symbolic % Use the symbolic class 'a' for all points
    ] table [x=HDI, y=DALY, meta=class] {
        DALY      HDI     class
        16308.69  0.586   a
        15646.25  0.586   a
        10894.78  0.66    a
        18959.8   0.569   a
        20523.52  0.519   a
        22231.99  0.565   a
        5480.14   0.747   a
        14693.67  0.608   a
        13577.7   0.639   a
        9211.46   0.667   a
        8879.85   0.648   a
        19505.74  0.566   a
        12332.61  0.609   a
        4516.86   0.718   a
        23272.98  0.54    a
        10871.48  0.649   a
        10613.26  0.696   a
        14740.89  0.634   a
        13725.8   0.693   a
        11674.19  0.68    a
        9511.21   0.708   a
        8964.62   0.662   a
        19687.96  0.552   a
        9938.81   0.638   a
        9279.86   0.653   a
        10593.49  0.615   a
        23622.08  0.536   a
        13544.05  0.632   a
        10680.65  0.575   a
    };
    \end{axis}
    \end{tikzpicture}
    \caption{Plot for the year 2011}
    \label{fig:daly2011}
\end{subfigure}

\caption{Comparison between the HDI and the DALY per 100,000 for communicable diseases in various Indian states in the years 2001 and 2011}
\label{fig:daly_comparison}
\end{figure}
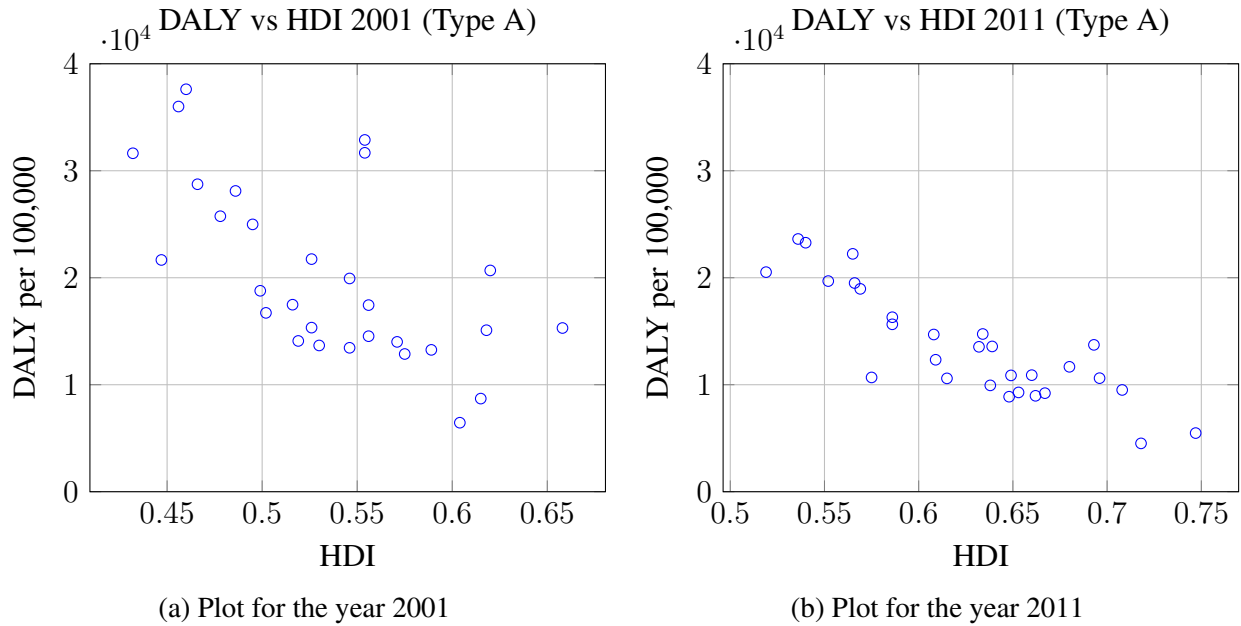

By comparing these two years, we can conclude three main takeaways:\\
Strengthening Inverse Relationship - From 2001 to 2011, the correlation between HDI and DALY became ‘less noisy’, and clearer and stronger. This shows how over the decade, improvements in human development, such as health, education, and living standards (all accounted for in HDI), was successfully associated with a reduction in the effect of communicable diseases across Indian states. \\
Overall decrease in Communicable Disease Burden (DALY values) - There was a notable decrease in values from 2001 to 2011. The peak DALY dropped from 38,000 to 24,000. This suggests that throughout the decade, changes were made that lessened the burden and effects of communicable diseases throughout India. This was likely due to improvements in health, sanitation, and disease control systems.\\
Progress in Human Development - The general downwards trend in the DALY values shows that there was overall advancement in human development across Indian states. \\

\textit{DALY type B - Noncommunicable diseases}\\
As displayed in Figure 2, the results are not intuitive and are different from the previous section of Type A DALY values.  

\begin{figure}[h!]
\centering % Center the entire figure

% --- First Subfigure (Left Plot) ---
\begin{subfigure}{0.49\textwidth}
    \centering
    \begin{tikzpicture}
    \begin{axis}[
        title={DALY vs HDI 2001 (Type B)},
        xlabel={HDI},
        ylabel={DALY per 100,000},
        grid=major,
        ymin=0,      % Sets the minimum y-axis value
        ymax=30000,   % Sets the maximum y-axis value
        scatter/classes={
            a={mark=o, fill=blue, draw=blue} % Define a class for blue circles
        }
    ]
    \addplot[
        scatter,
        only marks,
        scatter src=explicit symbolic % Use the symbolic class 'a' for all points
    ] table [x=HDI, y=DALY, meta=class] {
        DALY      HDI       class
        18393.83	0.495    a
        15614.77	0.478    a
        15614.77	0.499    a
        20651.36	0.486    a
        17815.51	0.432    a
        20343.33	0.554    a
        17541.23	0.615    a
        17742.68	0.526    a
        17353	    0.546    a
        18414.43	0.589    a
        15853.47	0.53     a
        18038.38	0.554    a
        19186.19	0.516    a
        20045.4	    0.604    a
        18671.37	0.456    a
        17972.09	0.556    a
        16103.29	0.556    a
        14927.35	0.447    a
        14726.97	0.571    a
        14070.85	0.519    a
        16673.01	0.658    a
        18539.59	0.575    a
        15342.09	0.466    a
        15651.88	0.546    a
        21438.9	    0.618    a
        18022.14	0.526    a
        17906.04	0.46     a
        21577.53	0.62     a
        18335.66	0.502    a
    };
    \end{axis}
    \end{tikzpicture}
    \caption{Plot for the year 2001}
    \label{fig:daly2001}
\end{subfigure}
\hfill % This creates a flexible space between the two subfigures
% --- Second Subfigure (Right Plot) ---
\begin{subfigure}{0.49\textwidth}
    \centering
    \begin{tikzpicture}
    \begin{axis}[
        title={DALY vs HDI 2011 (Type B)},
        xlabel={HDI},
        ylabel={DALY per 100,000},
        grid=major,
        ymin=0,      % Sets the minimum y-axis value
        ymax=30000,   % Sets the maximum y-axis value
        scatter/classes={
            a={mark=o, fill=blue, draw=blue} % Define a class for blue circles
        }
    ]
    \addplot[
        scatter,
        only marks,
        scatter src=explicit symbolic % Use the symbolic class 'a' for all points
    ] table [x=HDI, y=DALY, meta=class] {
        DALY        HDI     class
        19077.47	0.586    a
        18976.41	0.586    a
        15084.39	0.66     a
        20019.05	0.569    a
        16941.85	0.519    a
        21762.69	0.565    a
        19907.03	0.747    a
        19012.76	0.608    a
        19513.46	0.639    a
        19665.81	0.667    a
        16189.93	0.648    a
        16663.53	0.566    a
        20855.43	0.609    a
        22001.63	0.718    a
        18505.98	0.54     a
        18624.29	0.649    a
        17167.62	0.696    a
        15067.06	0.634    a
        16070.55	0.693    a
        15167.88	0.68     a
        17746.69	0.708    a
        20840.32	0.662    a
        16709.98	0.552    a
        16809.8	    0.638    a
        22914.46	0.653    a
        19140.91	0.615    a
        18316.04	0.536    a
        23869.98	0.632    a
        19258.02	0.575    a
    };
    \end{axis}
    \end{tikzpicture}
    \caption{Plot for the year 2011}
    \label{fig:daly2011}
\end{subfigure}

\caption{Comparison between the HDI and the DALY per 100,000 for noncommunicable diseases in various Indian states in the years 2001 and 2011}
\label{fig:daly_comparison}
\end{figure}
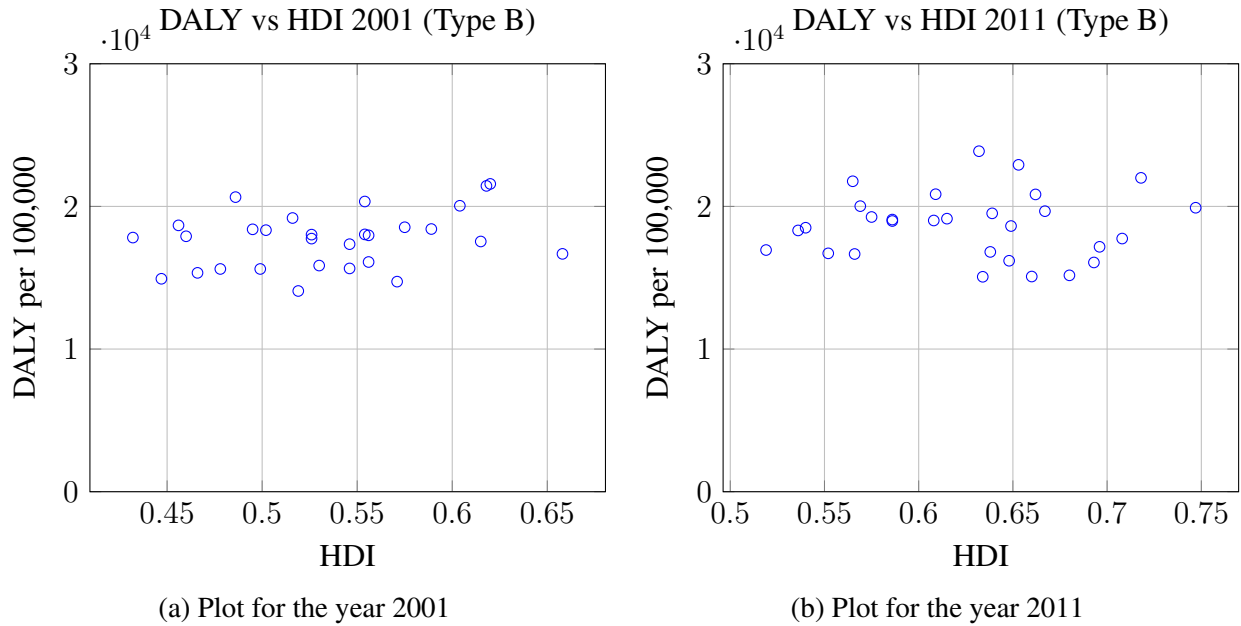

In both 2001 and 2011, we can see that the measure of the burden of noncommunicable diseases stayed fairly consistent, with most states being around the 15,000-22,000 area. Even with HDIs increasing through the decade, DALY values still stayed constant. This poses the question of whether noncommunicable diseases are significantly affected by the improvements in health and sanitation sectors during this period.\\

\textit{DALY Type C - Injuries}\\
Figure 3 displays the relationship between a state’s HDI index and their per 100,000 DALY value for injuries. Over the 10 years, the distribution of DALY due to injury changed notably. In 2001, the DALY values were widely dispersed, ranging from 2,500 to over 7,000 per 100,000 individuals. The data was highly variable with little indication of a relationship or trend between HDI and DALY. Some states with low HDI had high DALY, while some states with high HDI (0.61), also had high DALY. The lack of a clear trend suggests that, at the time, injury burden was influenced by factors not directly tied to overall development.

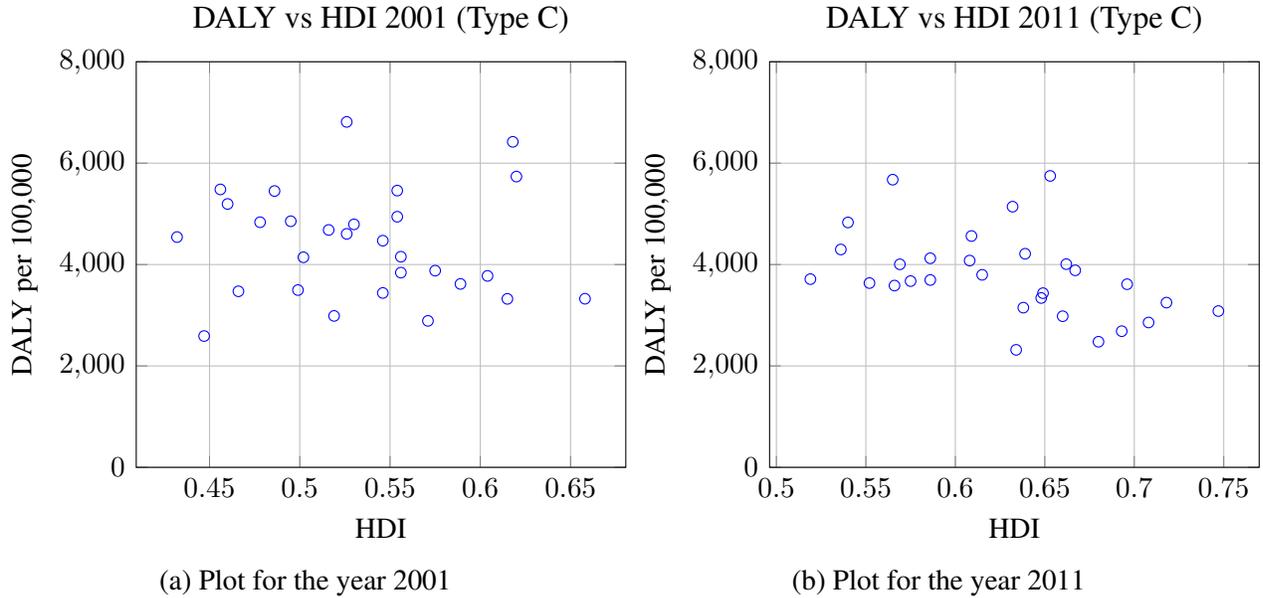
\begin{figure}[h!]
\centering % Center the entire figure

% --- First Subfigure (Left Plot) ---
\begin{subfigure}{0.49\textwidth}
    \centering
    \begin{tikzpicture}
    \begin{axis}[
        width=\linewidth, % <-- ADDED: Scales plot to fit the subfigure width
        title={DALY vs HDI 2001 (Type C)},
        xlabel={HDI},
        ylabel={DALY per 100,000},
        grid=major,
        ymin=0,      % Sets the minimum y-axis value
        ymax=8000,   % Sets the maximum y-axis value
        tick label style={font=\small}, % Optional: makes tick labels smaller
        label style={font=\small},      % Optional: makes axis labels smaller
        scatter/classes={
            a={mark=o, fill=blue, draw=blue} % Define a class for blue circles
        }
    ]
    \addplot[
        scatter,
        only marks,
        scatter src=explicit symbolic % Use the symbolic class 'a' for all points
    ] table [x=HDI, y=DALY, meta=class] {
        DALY      HDI      class
        4854.11   0.495    a
        4834.52   0.478    a
        3496.61   0.499    a
        5450.04   0.486    a
        4542.15   0.432    a
        5458.89   0.554    a
        3322.94   0.615    a
        6813.52   0.526    a
        4470.71   0.546    a
        3617.38   0.589    a
        4792.69   0.53     a
        4943.48   0.554    a
        4682.13   0.516    a
        3776.08   0.604    a
        5481.87   0.456    a
        4153.84   0.556    a
        3840.64   0.556    a
        2589.78   0.447    a
        2888.83   0.571    a
        2986.76   0.519    a
        3325.23   0.658    a
        3879.42   0.575    a
        3472.07   0.466    a
        3439.52   0.546    a
        6420.64   0.618    a
        4605.24   0.526    a
        5195.14   0.46     a
        5735.47   0.62     a
        4143.29   0.502    a
    };
    \end{axis}
    \end{tikzpicture}
    \caption{Plot for the year 2001}
    \label{fig:daly2001}
\end{subfigure}
\hfill % This creates a flexible space between the two subfigures
% --- Second Subfigure (Right Plot) ---
\begin{subfigure}{0.49\textwidth}
    \centering
    \begin{tikzpicture}
    \begin{axis}[
        width=\linewidth, % <-- ADDED: Scales plot to fit the subfigure width
        title={DALY vs HDI 2011 (Type C)},
        xlabel={HDI},
        ylabel={DALY per 100,000},
        grid=major,
        ymin=0,      % Sets the minimum y-axis value
        ymax=8000,   % Sets the maximum y-axis value
        tick label style={font=\small}, % Optional: makes tick labels smaller
        label style={font=\small},      % Optional: makes axis labels smaller
        scatter/classes={
            a={mark=o, fill=blue, draw=blue} % Define a class for blue circles
        }
    ]
    \addplot[
        scatter,
        only marks,
        scatter src=explicit symbolic % Use the symbolic class 'a' for all points
    ] table [x=HDI, y=DALY, meta=class] {
        DALY      HDI      class
        4123.92   0.586    a
        3695.19   0.586    a
        2980.51   0.66     a 
        4004.6    0.569    a
        3712.14   0.519    a
        5674.1    0.565    a
        3082.54   0.747    a
        4078.23   0.608    a
        4212.86   0.639    a
        3885.63   0.667    a
        3340.07   0.648    a
        3584.96   0.566    a
        4562.52   0.609    a
        3249.04   0.718    a
        4830.66   0.54     a
        3435.36   0.649    a
        3611.84   0.696    a
        2317.44   0.634    a
        2685.35   0.693    a
        2477.2    0.68     a
        2857.11   0.708    a
        4007.73   0.662    a
        3634.52   0.552    a
        3149.65   0.638    a
        5748.17   0.653    a
        3797.24   0.615    a
        4297.29   0.536    a
        5141.56   0.632    a
        3673.62   0.575    a
    };
    \end{axis}
    \end{tikzpicture}
    \caption{Plot for the year 2011}
    \label{fig:daly2011}
\end{subfigure}

\caption{Comparison between the HDI and the DALY per 100,000 for injuries in various Indian states in the years 2001 and 2011}
\label{fig:daly_comparison}
\end{figure}

But by 2011, DALY values became more tightly clustered, with most states having values between 2,500 and 5,500 DALYs per 100,000. This narrowing range indicates a more consistent outcome between HDI levels and a general reduction in injury burden. In addition, there is a slight negative correlation between HDI and DALY. Higher HDI states have lower DALY values. This indicates that improvements in healthcare systems and infrastructure led to fewer disability-causing injuries and more effective treatments. \\

\subsubsection{Gender Ratio study}

Analyzing Indian census and health data displayed the gender disparities in disability. In nearly all Indian states, the disabled male–female ratio was consistently higher than the overall gender ratio (shown in Figure 4), showing how men are overrepresented in the disabled population relative to their proportion in the total population. In India in 2001, there were 12.6 million men with disabilities compared to 9.4 million women. This is a disabled male–female ratio of ~1.34, while the overall male–female ratio was just \~1.06.

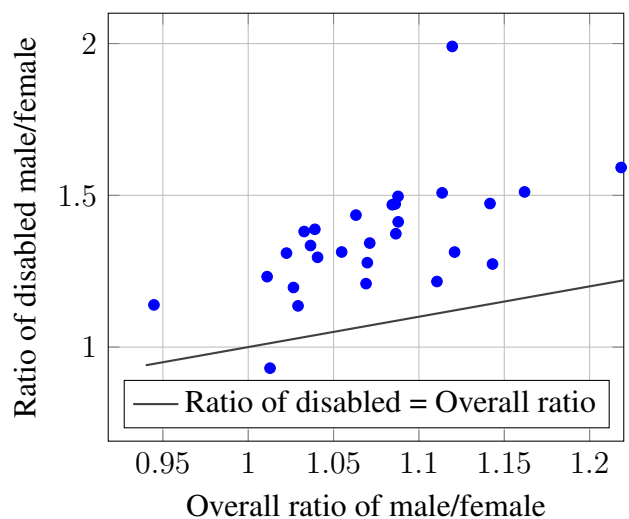
\begin{figure}[h!]
\centering
\begin{tikzpicture}
\begin{axis}[
    title={2001 Ratio of disabled male/female vs. Ratio of male/female},
    xlabel={Overall ratio of male/female},
    ylabel={Ratio of disabled male/female},
    grid=major,
    legend pos=south west,
    xmin=0.918, xmax=1.22,
    ymin=0.69, ymax=2.1,
    scatter/classes={ 
        a={mark=*, blue} 
    }
]

% --- Scatter Plot ---
\addplot[
    scatter,
    only marks,
    scatter src=explicit symbolic,
    forget plot % <-- ADDED: This excludes the scatter points from the legend.
] table [x=x_val, y=y_val, meta=class] { % Tell the table to use the 'meta' column
    x_val         y_val         class % Add the 'class' column back in
    1.071918856   1.355279367   a
    1.022419343   1.309573442   a
    1.119518867   1.990574506   a
    1.069771062   1.278077531   a
    1.087765214   1.496559249   a
    1.011065405   1.232028663   a
    1.040622634   1.295772595   a
    1.086477206   1.373354431   a
    1.161885796   1.511216702   a
    1.032769669   1.380697951   a
    1.120882149   1.313035903   a
    1.063108347   1.434873037   a
    1.036501503   1.334605734   a
    0.944777423   1.138916222   a
    1.087851107   1.41254464    a
    1.08439611    1.469002619   a
    1.026510145   1.19628483    a
    1.029186119   1.135770429   a
    1.069027905   1.209023179   a
    1.1105595     1.216006021   a
    1.218468902   1.591755115   a
    1.141647399	  1.472944713   a
    1.086123187	  1.47139389    a
    1.143113006	  1.273610181   a
    1.012776711	  0.9305052115  a   
    1.054751576	  1.313277601   a
    1.113602023	  1.508139142   a
    1.039030125	  1.388045611   a
    1.071189819	  1.342675675   a
};

% --- 1-to-1 Trendline ---
\addplot[
    domain=0.94:1.22,
    samples=2,
    thick,
    darkgray,
] {x};
\addlegendentry{Ratio of disabled = Overall ratio} % This now correctly refers to the line plot

\end{axis}
\end{tikzpicture}
\caption{Scatterplot comparing the ratio of disabled males to females (y-axis) against the overall male-to-female population ratio (x-axis) for 2001. The line indicates a 1:1 correspondence}
\label{fig:ratio_comparison}
\end{figure}

Arunachal Pradesh is an outlier in this data, having a disabled male-female ratio of nearly 2. Census Data showed that 66.6\% of persons with disabilities in the state were male, the highest proportion in India. We can likely explain this outlier data through a few reasons. Firstly, Arunachal Pradesh has a population of only 1.10 million. This is a number smaller than other Indian regions. Because it has a smaller population, it is more sensitive to small numerical changes and the overall gender ratio is more volatile. This, combined with the stigma around female disability, which causes underreporting and skewed data, makes the final disability ratio differ from the expected value.  Secondly, males in this region often work in physically demanding and more risky jobs, raising the likelihood of a disability-causing injury. Overall, this outlier data is likely due to smaller population size, reporting disparities, and male-dominant labor.

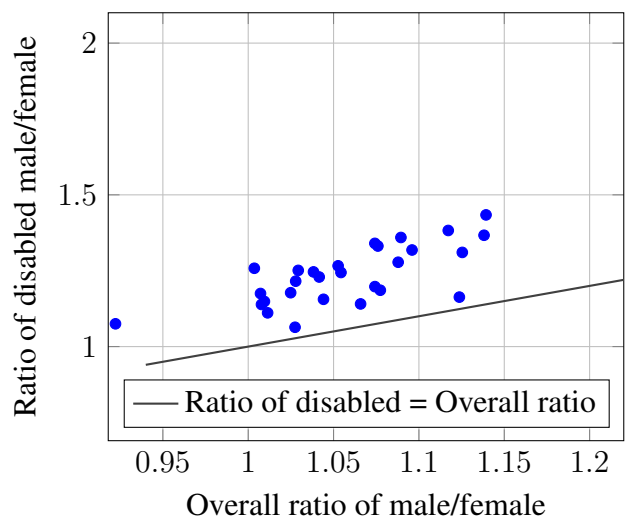
\begin{figure}[h!]
\centering
\begin{tikzpicture}
\begin{axis}[
    title={2011 Ratio of disabled male/female vs. Ratio of male/female},
    xlabel={Overall ratio of male/female},
    ylabel={Ratio of disabled male/female},
    grid=major,
    legend pos=south west,
    xmin=0.918, xmax=1.22,
    ymin=0.69, ymax=2.1,
    scatter/classes={ 
        a={mark=*, blue} 
    }
]

% --- Scatter Plot ---
\addplot[
    scatter,
    only marks,
    scatter src=explicit symbolic,
    forget plot % <-- ADDED: This excludes the scatter points from the legend.
] table [x=x_val, y=y_val, meta=class] { % Tell the table to use the 'meta' column
    x_val         y_val         class % Add the 'class' column back in
    1.063524352	  1.267384135     a
    1.007202773	  1.174937725     a
    1.065834596	  1.140603731     a
    1.044104817   1.155851446     a
    1.089456968	  1.359538176     a
    1.009486243   1.14870171      a
    1.027432392   1.063765941     a
    1.087839922	  1.278011587     a
    1.138149918   1.366884566     a
    1.02930888	  1.251119646     a
    1.125413853	  1.310358945     a
    1.054334652	  1.244159205     a
    1.027814631	  1.215560396     a
    0.9222472932  1.075091859     a
    1.0741922	  1.340135408     a
    1.075959394   1.3313474       a
    1.00777367	  1.13885946      a
    1.011372442	  1.11123815      a
    1.024862189	  1.177535191     a
    1.07422108	  1.197656308     a
    1.139310939	  1.433934696     a
    1.117186117	  1.38263901      a
    1.077386518	  1.185740425     a
    1.12369438	  1.163058991     a
    1.003580211	  1.258107914     a
    1.041585604	  1.229282151     a
    1.095966677	  1.31830386      a
    1.038244574	  1.246129599     a
    1.052666795	  1.266175405     a
};

% --- 1-to-1 Trendline ---
\addplot[
    domain=0.94:1.22,
    samples=2,
    thick,
    darkgray,
] {x};
\addlegendentry{Ratio of disabled = Overall ratio} % This now correctly refers to the line plot

\end{axis}
\end{tikzpicture}
\caption{Scatterplot comparing the ratio of disabled males to females (y-axis) against the overall male-to-female population ratio (x-axis) for 2011. The line indicates a 1:1 correspondence}
\label{fig:ratio_comparison}
\end{figure}

As shown in Figure 5, in 2011, while there is still a general trend of male overrepresentation among disabled individuals, the points are more clustered and there are fewer extreme outliers. This suggests a decrease in risks contributing to higher male disability rates and also aligns with other DALY data that show declines in overall injury-related and disease burdens among men over this time period. However, the overrepresentation in male disability is still just as evident as in 2001, indicating that changes to address these disparities has been slow. \\
The patterns and trends from 2001 to 2011, such as how communicable diseases DALYs dropped as HDI values rose, can be used to predict future values for 2031 through machine learning algorithms.

\subsection{Current Trend (2021)}
\subsubsection{DALY vs HDI}

In 2021, states with higher HDI values (above 0.70) consistently show the lowest DALYs, while states with lower HDI values still exhibit a higher burden of communicable diseases.
\begin{figure}[h!]
\centering
\begin{tikzpicture}
\begin{axis}[
    title={DALY vs HDI 2021 (Type A)},
    xlabel={HDI},
    ylabel={DALY per 100,000},
    grid=major,
    ymin=0,      % Sets the minimum y-axis value
    ymax=40000,   % Sets the maximum y-axis value
    scatter/classes={
        a={mark=o, fill=blue, draw=blue} % Define a class for blue circles
    }
]
\addplot[
    scatter,
    only marks,
    scatter src=explicit symbolic % Use the symbolic class 'a' for all points
] table [x=HDI, y=DALY, meta=class] {
    DALY         HDI       class
    12450.01	0.633      a
    10929.28	0.630      a
    8104.54	    0.670      a
    13574.45	0.603      a
    13347.01	0.566      a
    16489.04	0.613      a
    8843.12	    0.747      a
    11363.43	0.635      a
    11301.25	0.683      a
    8767.91	    0.703      a
    7676.11	    0.706      a
    12037.84	0.588      a
    10377.95	0.661      a
    7823.63	    0.745      a
    14867.36	0.599      a
    9888.82	    0.683      a
    11252.62	0.671      a
    11100.05	0.637      a
    10485.95	0.697      a
    9244.34	    0.667      a
    7690.56	    0.722      a
    8556.34	    0.685      a
    13107.22	0.640      a
    9604.47	    0.699      a
    9378.04	    0.679      a
    9421.25	    0.612      a
    17940.31	0.598      a
    12495.32	0.669      a
    8057.91	    0.623      a
};
\end{axis}
\end{tikzpicture}
\caption{Scatterplot of Disability-Adjusted Life Years (DALY) per 100,000 for Type A Communicable Diseases against the Human Development Index (HDI) for 2021.}
\end{figure}
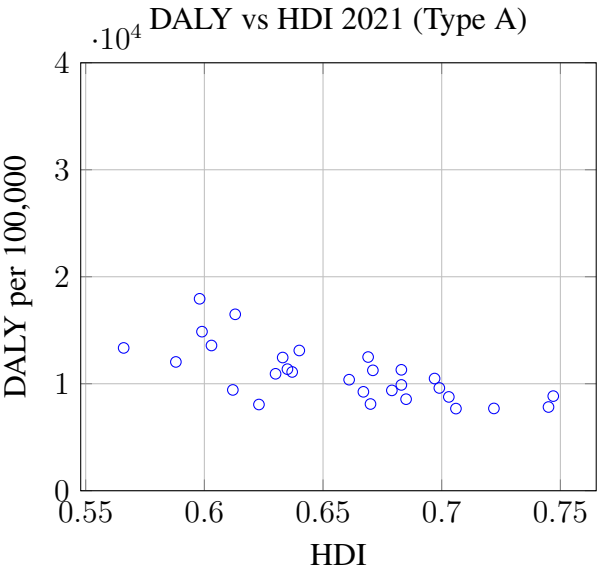

The maximum DALY has further decreased to less than 18,000 from \~24,000 in 2011, and the entire cluster of data points has shifted to the right on the HDI axis and down on the DALY axis. This indicates continued progress in both variables. The lowest HDI scores in 2021 are now comparable to the mid-range scores of the previous decade. In improvements in HDI and DALY demonstrate broad and ongoing improvements in education, living standards, and health infrastructure across all states.\\

\begin{figure}[h!]
\centering
\begin{tikzpicture}
\begin{axis}[
    title={DALY vs HDI 2021 (Type B)},
    xlabel={HDI},
    ylabel={DALY per 100,000},
    grid=major,
    ymin=0,      % Sets the minimum y-axis value
    ymax=30000,   % Sets the maximum y-axis value
    scatter/classes={
        a={mark=o, fill=blue, draw=blue} % Define a class for blue circles
    }
]
\addplot[
    scatter,
    only marks,
    scatter src=explicit symbolic % Use the symbolic class 'a' for all points
] table [x=HDI, y=DALY, meta=class] {
    DALY         HDI       class
    20465.66	0.633      a
    20959.98	0.630      a
    15449.55	0.670      a
    19353.23	0.603      a
    16411.07	0.566      a
    21651.49	0.613      a
    22860.01	0.747      a
    21905.88	0.635      a
    20182.21	0.683      a
    22175.53	0.703      a
    18141.66	0.706      a
    15200.73	0.588      a
    22478.15	0.661      a
    25105.18	0.745      a
    19625.54	0.599      a
    21402.28	0.683      a
    18855.69	0.671      a
    15805.02	0.637      a
    17228.3	    0.697      a
    16736.27	0.667      a
    18387.79	0.722      a
    23200.49	0.685      a
    18265.23	0.640      a
    18101.99	0.699      a
    24709.25	0.679      a
    20563.25	0.612      a
    19808.38	0.598      a
    24772.73	0.669      a
    21673.08	0.623      a
};
\end{axis}
\end{tikzpicture}
\caption{Scatterplot of Disability-Adjusted Life Years (DALY) per 100,000 for Type B Noncommunicable Diseases against the Human Development Index (HDI) for 2021.}
\end{figure}
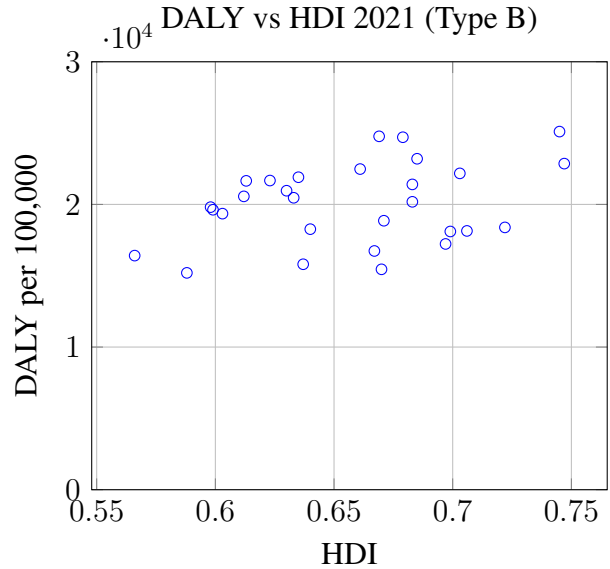

Figure 7 supports the observations from Figure 2. Despite continued and significant increases in HDI across all states, in 2021, the overall noncommunicable disease burden appears to have slightly increased from 2011. Many states are now clustering in the 18,000-25,000 DALY range. Additionally, a subtle positive correlation is now visible. States with higher HDI now tend to have a slightly higher burden of noncommunicable disease.\\
This observed shift is an example of an epidemiological transition. As a nation improves sanitation and basic healthcare, the burden of communicable diseases drops, but the burden of noncommunicable diseases increase. This is due to development leading to new challenges. As healthcare improves, people live longer and chronic conditions to develop. Economic growth is also often linked to lifestyle shifts. People begin consuming processed foods, work in sedentary office jobs, etc. All of these are risk factors for noncommunicable diseases. The 2021 plot visually captures this transition and displays a turning point for public health policy toward managing these chronic conditions caused by modernization.

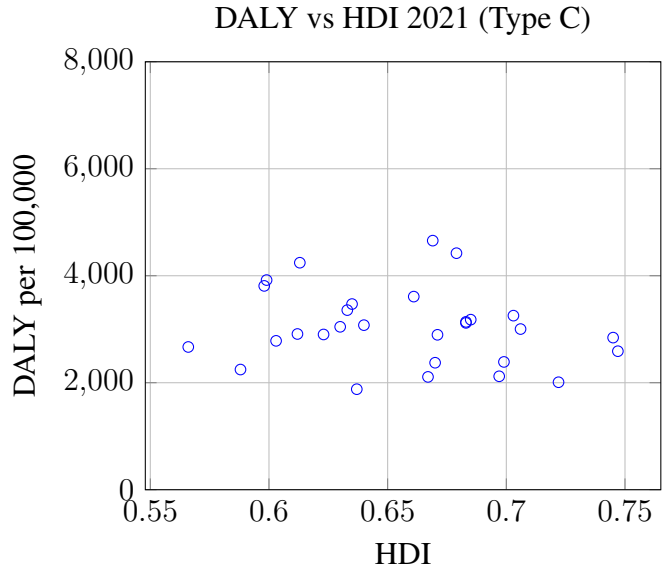
\begin{figure}[h!]
\centering
\begin{tikzpicture}
\begin{axis}[
    title={DALY vs HDI 2021 (Type C)},
    xlabel={HDI},
    ylabel={DALY per 100,000},
    grid=major,
    ymin=0,      % Sets the minimum y-axis value
    ymax=8000,   % Sets the maximum y-axis value
    scatter/classes={
        a={mark=o, fill=blue, draw=blue} % Define a class for blue circles
    }
]
\addplot[
    scatter,
    only marks,
    scatter src=explicit symbolic % Use the symbolic class 'a' for all points
] table [x=HDI, y=DALY, meta=class] {
    DALY         HDI       class
    3357.39	    0.633       a
    3045.13	    0.630       a
    2373.71	    0.670       a
    2782.87	    0.603       a
    2668.25	    0.566       a
    4243.35	    0.613       a
    2589.28	    0.747       a
    3470.66	    0.635       a
    3118.47	    0.683       a
    3253.94	    0.703       a
    3003.21	    0.706       a
    2246.45	    0.588       a
    3609.9	    0.661       a
    2843.31	    0.745       a
    3918.99	    0.599       a
    3137.4	    0.683       a
    2895.74	    0.671       a
    1879.03	    0.637       a
    2120.11	    0.697       a
    2108.92	    0.667       a
    2009.04	    0.722       a
    3179.04	    0.685       a
    3076.25	    0.640       a
    2388.69	    0.699       a
    4421.54	    0.679       a
    2910.95	    0.612       a
    3810.04	    0.598       a
    4656.13	    0.669       a
    2901.97	    0.623       a
};
\end{axis}
\end{tikzpicture}
\caption{Scatterplot of Disability-Adjusted Life Years (DALY) per 100,000 for Type C Injuries against the Human Development Index (HDI) for 2021.}
\end{figure}
Figure 8 displays how, by 2021, overall injury burden continued to increase, with DALY values becoming even more tightly clustered than in 2011 and now mostly ranging between 2,000 and 4,500 per 100,000. This indicates continued success in decreasing the most severe outcomes of injuries across the country. However, in 2021, there is no longer the slight negative correlation that was emerging in 2011. In figure 8, there is now significant variability without a clear linear trend, similar to the 2001 (Figure 3) distribution but at a lower DALY level.\\
This suggests more complex relationships. While modernization reduces the overall injury burden nationwide, the types of injury risks occurring may be changing. For example, as states develop, traffic accidents or other occupational injuries may increase, preventing a negative correlation from occurring. The injury burden in 2021 appears to be influenced by a set of factors not captured by the HDI metric.

\subsubsection{Gender proportions}
The 2021 Census of India was postponed indefinitely due to the COVID-19 pandemic. As of late 2025, it has not yet been taken up. Since the national census is the most reliable source for demographic data, including population counts by gender, a gender proportion study is not undertaken for 2021.

\subsection{Future trend (2031) and four-decade projection analysis }
As explained in the Methodology section above, the prediction for HDI, noncommunicable DALY, and Injury DALY utilized a Linear Regression model, while Communicable was done with an Exponential Decay model. The 2031 HDI value was first predicted, and then passed in as a variable to aid in the DALY predictions.  \\
A few representative predicted data are shown in Table 1 below:

\begin{table}[h!]
\centering
\caption{Predicted DALY and HDI values for various Indian states for the year 2031. The full dataset can be found at \url{https://github.com/Arunimad/India_DALY_HDI_prediction.git}}
\label{tab:predicted_data_2031}

% Use \small font size to help the table fit better
\small 

% Increase vertical spacing in rows for better readability
\renewcommand{\arraystretch}{1.5} 

\begin{tabular}{|l|p{1.8cm}|p{2cm}|p{1.8cm}|p{1.7cm}|} 
\hline
% --- Header Row ---
\textbf{Area Name} & 
\textbf{Comm DALY \newline (Predicted)} & 
\textbf{Noncomm DALY \newline (Predicted)} & 
\textbf{Injury DALY \newline (Predicted)} & 
\textbf{HDI \newline (Predicted)} \\ 
\hline

% --- Data Rows ---
INDIA & 8633.95 & 20369.20 & 2671.37 & 0.709 \\ \hline
ANDHRA PRADESH & 7151.02 & 23758.10 & 2100.05 & 0.717 \\ \hline
ARUNACHAL PRADESH & 5723.95 & 15041.89 & 2078.52 & 0.781 \\ \hline
ASSAM & 9627.15 & 20347.20 & 1522.56 & 0.670 \\ \hline
BIHAR & 8864.07 & 16760.08 & 1837.19 & 0.640 \\ \hline
\end{tabular}
\end{table}

\subsubsection{Analysis of HDI trends over four decades}
Figure 9 displays HDI data for the years 2001 (blue), 2011 (red), 2021 (orange), and the model’s projection for 2031 (teal). It displays a universal and significant continuation of positive HDI trends across all states, with national HDI reaching 0.709 by 2031. This increase reflects the assumption that widespread advancements in areas surrounding living standards, education, and health will be made over the next decade. The plot also displays development levels across India. States such as Goa and Kerala have higher predicted 2031 HDI values, while Madhya Pradesh and Jharkhand are situated lower on the y-axis.

% --- Data Table for HDI Values ---
% Data extracted from the provided PDF 
\pgfplotstableread[col sep=comma]{
Area,y2001,y2011,y2021,y2031
INDIA,0.495,0.586,0.633,0.709
ANDHRA PRADESH,0.478,0.586,0.630,0.717
ARUNACHAL PRADESH,0.499,0.660,0.670,0.781
ASSAM,0.486,0.569,0.603,0.670
BIHAR,0.432,0.519,0.566,0.640
CHHATTISGARH,0.554,0.565,0.613,0.636
GOA,0.615,0.747,0.747,0.835
GUJARAT,0.526,0.608,0.635,0.699
HARYANA,0.546,0.639,0.683,0.760
HIMACHAL PRADESH,0.589,0.667,0.703,0.767
JAMMU \& KASHMIR,0.530,0.648,0.706,0.804
JHARKHAND,0.554,0.566,0.588,0.603
KARNATAKA,0.516,0.609,0.661,0.740
KERALA,0.604,0.718,0.745,0.830
MADHYA PRADESH,0.456,0.540,0.599,0.675
MAHARASHTRA,0.556,0.649,0.683,0.756
MANIPUR,0.556,0.696,0.671,0.756
MEGHALAYA,0.447,0.634,0.637,0.763
MIZORAM,0.571,0.693,0.697,0.780
NAGALAND,0.519,0.680,0.667,0.770
NCT OF DELHI,0.658,0.708,0.722,0.760
PUNJAB,0.575,0.662,0.685,0.751
RAJASTHAN,0.466,0.552,0.640,0.727
SIKKIM,0.546,0.638,0.699,0.781
TAMIL NADU,0.618,0.653,0.679,0.711
TRIPURA,0.526,0.615,0.612,0.670
UTTAR PRADESH,0.460,0.536,0.598,0.669
UTTARAKHAND,0.620,0.632,0.669,0.689
WEST BENGAL,0.502,0.575,0.623,0.688
}\HDIdata

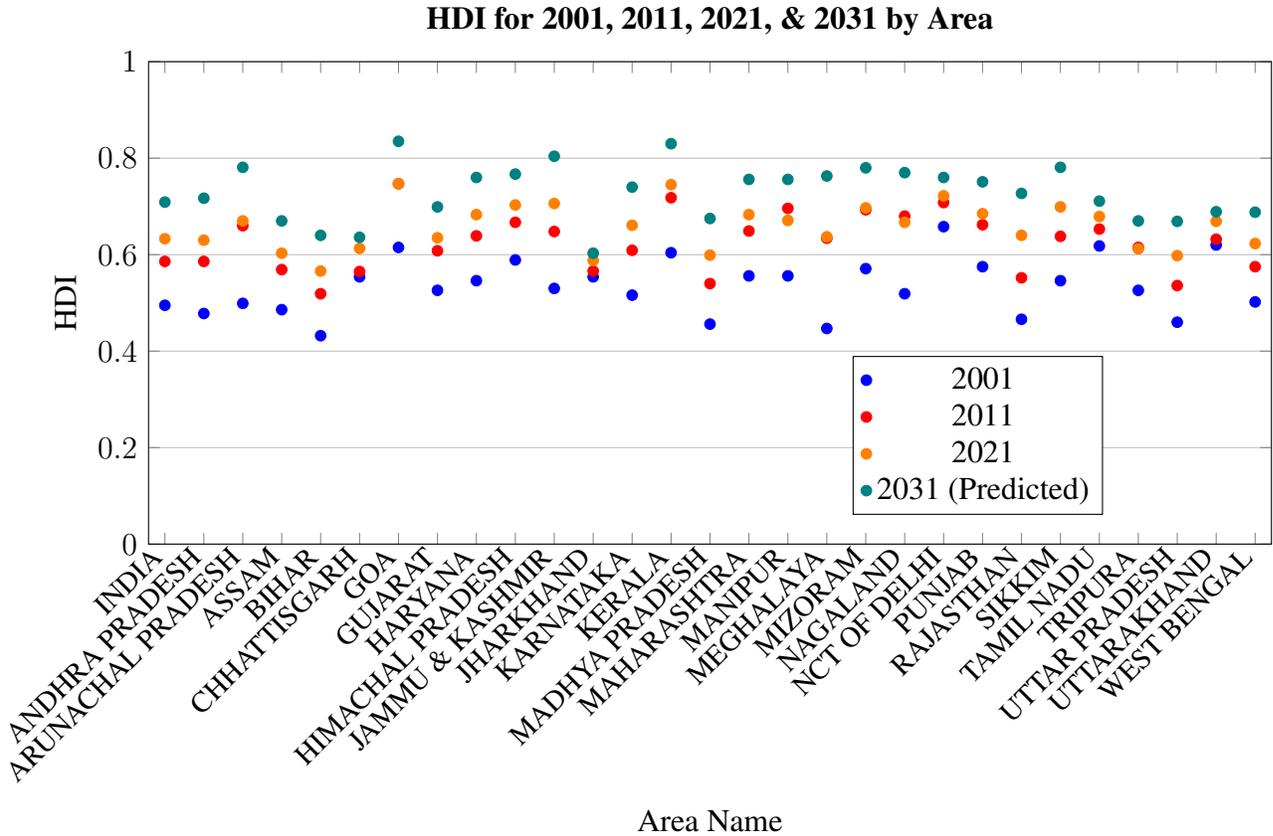
\begin{figure}[h!]
\centering
\begin{tikzpicture}
\begin{axis}[
    % Set the plot title based on the PDF [cite: 1]
    title={\textbf{HDI for 2001, 2011, 2021, \& 2031 by Area}},
    xlabel={Area Name},
    ylabel={HDI},
    % Define all area names for the symbolic x-axis
    symbolic x coords={
        INDIA, ANDHRA PRADESH, ARUNACHAL PRADESH, ASSAM, BIHAR, CHHATTISGARH, 
        GOA, GUJARAT, HARYANA, HIMACHAL PRADESH, JAMMU \& KASHMIR, JHARKHAND, 
        KARNATAKA, KERALA, MADHYA PRADESH, MAHARASHTRA, MANIPUR, MEGHALAYA, 
        MIZORAM, NAGALAND, NCT OF DELHI, PUNJAB, RAJASTHAN, SIKKIM, TAMIL NADU, 
        TRIPURA, UTTAR PRADESH, UTTARAKHAND, WEST BENGAL
    },
    xtick=data, % Place ticks at each x-coordinate
    xticklabel style={rotate=45, anchor=east, font=\small}, % Rotate labels for readability
    ymin=0,
    ymax=1.0, % Set y-axis for HDI scale (0 to 1) [cite: 5, 11]
    enlarge x limits=0.015,
    ymajorgrids=true, % Add horizontal grid lines
    width=\textwidth,
    height=8cm,
    % Configure the legend to be above the plot and horizontal [cite: 1]
    legend style={
        at={(0.85, 0.39)},      % Position at the top-right corner of the plot area
        anchor=north east,     % Anchor the legend's top-right corner to that point
        legend columns=1       % Use a single column for a vertical layout
    }
]

% Plot the data for each year with distinct colors
\addplot[only marks, mark=*, mark options={fill=blue, draw=blue}] 
    table [x=Area, y=y2001] {\HDIdata};
\addlegendentry{2001}

\addplot[only marks, mark=*, mark options={fill=red, draw=red}] 
    table [x=Area, y=y2011] {\HDIdata};
\addlegendentry{2011}

\addplot[only marks, mark=*, mark options={fill=orange, draw=orange}] 
    table [x=Area, y=y2021] {\HDIdata};
\addlegendentry{2021}

\addplot[only marks, mark=*, mark options={fill=teal, draw=teal}] 
    table [x=Area, y=y2031] {\HDIdata};
\addlegendentry{2031 (Predicted)}

\end{axis}
\end{tikzpicture}
\caption{Scatterplot showing the HDI across various areas in India for the years 2001, 2011, 2021, and with predictions for 2031.}
\end{figure}

\subsubsection{Analysis of DALY (Type A - Communicable) trends over four decades}
Figure 10 shows the Communicable DALYs for the same four years by area. 
Figure 10 shows the Communicable DALYs for the same four years by area. The spacing of the dots shows the steep and consistent downward trajectory of DALYs for nearly every state, with the blue (2001) dots being high on the y-axis while the teal (2031) dots having extremely low predicted values. The decreasing space between the dots shows the principle of diminishing returns, as shown by the distance between 2001 (blue) and 2011 (red) being noticeably larger than the distance between 2011 to 2021 (orange). This slowing rate is due to initial, high-impact interventions having already been completed, and is a justification of the exponential decay model selected.

This plot also allows for comparison of communicable DALY burdens among states in 2031:\\
High-Burden States (DALYs > 10,000): Uttar Pradesh, Mizoram, Uttarakhand, Chhattisgarh\\
Lower-Burden States (DALYs < 6,000): West Bengal, NCT of Delhi, Jammu \& Kashmir, Goa, Arunachal Pradesh

\pgfplotstableread[col sep=comma]{
Area,y2001,y2011,y2021,y2031
INDIA,24986.39,16308.69,12450.01,8633.95
ANDHRA PRADESH,25749.53,15646.25,10929.28,7151.02
ARUNACHAL PRADESH,18780.06,10894.78,8104.54,5723.95
ASSAM,28111.71,18959.8,13574.45,9627.15
BIHAR,31633.82,20523.52,13347.01,8864.07
CHHATTISGARH,32879.71,22231.99,16489.04,12713.11
GOA,8703.15,5480.14,8843.12,5998.56
GUJARAT,21741.16,14693.67,11363.43,8260.65
HARYANA,19931.57,13577.7,11301.25,8221.39
HIMACHAL PRADESH,13257.39,9211.46,8767.91,6638.10
JAMMU \& KASHMIR,13669.25,8879.85,7676.11,5431.67
JHARKHAND,31677.04,19505.74,12037.84,7694.16
KARNATAKA,17480.04,12332.61,10377.95,7753.77
KERALA,6449.2,4516.86,7823.63,6146.10
MADHYA PRADESH,36002.52,23272.98,14867.36,9645.51
MAHARASHTRA,17442.08,10871.48,9888.82,6887.26
MANIPUR,14535.66,10613.26,11252.62,9288.89
MEGHALAYA,21661.84,14740.89,11100.05,8927.80
MIZORAM,14004.55,13725.8,10485.95,10485.95
NAGALAND,14091.6,11674.19,9244.34,8721.95
NCT OF DELHI,15300.31,9511.21,7690.56,5329.55
PUNJAB,12871.92,8964.62,8556.34,6521.91
RAJASTHAN,28739.27,19687.96,13107.22,8879.97
SIKKIM,13451.64,9938.81,9604.47,7651.89
TAMIL NADU,15095.22,9279.86,9378.04,6663.76
TRIPURA,15336.77,10593.49,9421.25,7599.92
UTTAR PRADESH,37617.04,23622.08,17940.31,11978.07
UTTARAKHAND,20681.1,13544.05,12495.32,10104.21
WEST BENGAL,16716.87,10680.65,8057.91,5424.96
}\DALYdata

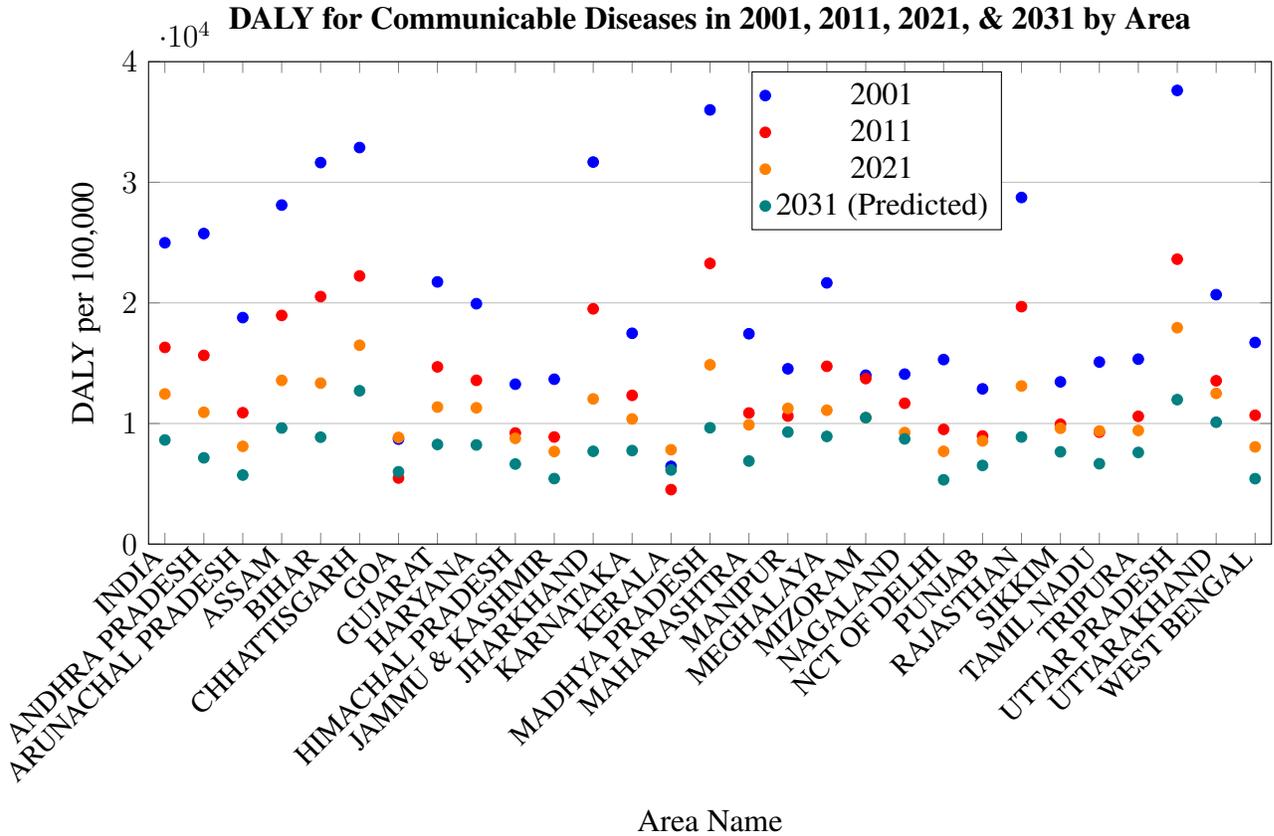
\begin{figure}[h!]
\centering
\begin{tikzpicture}
\begin{axis}[
    title={\textbf{DALY for Communicable Diseases in 2001, 2011, 2021, \& 2031 by Area}},
    xlabel={Area Name},
    ylabel={DALY per 100,000},
    symbolic x coords={
        INDIA, ANDHRA PRADESH, ARUNACHAL PRADESH, ASSAM, BIHAR, CHHATTISGARH, 
        GOA, GUJARAT, HARYANA, HIMACHAL PRADESH, JAMMU \& KASHMIR, JHARKHAND, 
        KARNATAKA, KERALA, MADHYA PRADESH, MAHARASHTRA, MANIPUR, MEGHALAYA, 
        MIZORAM, NAGALAND, NCT OF DELHI, PUNJAB, RAJASTHAN, SIKKIM, TAMIL NADU, 
        TRIPURA, UTTAR PRADESH, UTTARAKHAND, WEST BENGAL
    },
    xtick=data,
    xticklabel style={rotate=45, anchor=east, font=\small},
    ymin=0,
    ymax=40000,
    enlarge x limits=0.015,
    ymajorgrids=true,
    width=\textwidth,
    height=8cm,
    legend style={
        at={(0.76, 0.98)},      % Position at the top-right corner of the plot area
        anchor=north east,     % Anchor the legend's top-right corner to that point
        legend columns=1       % Use a single column for a vertical layout
    }
]

% --- Plot each year ---
% By removing 'scatter', we disable the default colormap behavior
% and allow the 'mark options' to take full control of the color.

\addplot[only marks, mark=*, mark options={fill=blue, draw=blue}] 
    table [x=Area, y=y2001] {\DALYdata};
\addlegendentry{2001}

\addplot[only marks, mark=*, mark options={fill=red, draw=red}] 
    table [x=Area, y=y2011] {\DALYdata};
\addlegendentry{2011}

\addplot[only marks, mark=*, mark options={fill=orange, draw=orange}] 
    table [x=Area, y=y2021] {\DALYdata};
\addlegendentry{2021}

\addplot[only marks, mark=*, mark options={fill=teal, draw=teal}] 
    table [x=Area, y=y2031] {\DALYdata};
\addlegendentry{2031 (Predicted)}

\end{axis}
\end{tikzpicture}
\caption{Scatterplot showing DALY per 100,000 for communicable diseases across various areas in India for the years 2001, 2011, 2021, and with predictions for 2031.}
\end{figure}

\subsubsection{Analysis of DALY (Type B - Noncommunicable) trends over four decades}

Figure 11 displays the Noncommunicable DALYs for the same four years by area. 
Unlike the clear, uniform decline observed for communicable diseases, the historical trend for noncommunicable diseases (NCD) varies significantly between states. For some regions (e.g., Kerala, Punjab) the data points show an upward trend, displaying that disease burdens increase by decade. But for others (e.g., Gujarat, Goa) the trend is unpredictable, with initial increases followed by decreases. This lack of a clear trend displays the challenges in modeling noncommunicable diseases and supports the decision to use a stable predictive model.

Despite volatility, the model’s 2031 projections (teal) dots are frequently positioned higher than the preceding data points. This indicates that the model generally predicts a continued increase in the NCD burden for the majority of states. These results reflect the linear model’s ability to capture the positive correlations between rising HDI and the increased prevalence of chronic, noncommunicable diseases. The model's results are also consistent with the epidemiological public health theory.

This plot also allows for comparison of noncommunicable DALY burdens among states in 2031:\\
Higher-Burden States (DALYs > 25000): Kerala, Punjab, Uttarakhand \\
Lower-Burden States (DALYs < 16000): Arunachal Pradesh, Jharkhand, Meghalaya, Nagaland 

% --- Data Table for Noncommunicable DALY Values ---
% Data extracted from the provided PDF
\pgfplotstableread[col sep=comma]{
Area,y2001,y2011,y2021,y2031
INDIA,18393.83,19077.47,20465.66,20369.20
ANDHRA PRADESH,15614.77,18976.41,20959.98,23758.10
ARUNACHAL PRADESH,15614.77,15084.39,15449.55,15041.89
ASSAM,20651.36,20019.05,19353.23,20347.20
BIHAR,17815.51,16941.85,16411.07,16760.08
CHHATTISGARH,20343.33,21762.69,21651.49,22121.25
GOA,17541.23,19907.03,22860.01,20396.35
GUJARAT,17742.68,19012.76,21905.88,19476.81
HARYANA,17353.00,19513.46,20182.21,19087.25
HIMACHAL PRADESH,18414.43,19665.81,22175.53,21897.42
JAMMU \& KASHMIR,15853.47,16189.93,18141.66,18320.16
JHARKHAND,18038.38,16663.53,15200.73,14828.87
KARNATAKA,19186.19,20855.43,22478.15,24043.32
KERALA,20045.40,22001.63,25105.18,26664.65
MADHYA PRADESH,18671.37,18505.98,19625.54,19806.75
MAHARASHTRA,17972.09,18624.29,21402.28,22218.18
MANIPUR,16103.29,17167.62,18855.69,18830.93
MEGHALAYA,14927.35,15067.06,15805.02,15788.91
MIZORAM,14726.97,16070.55,17228.30,17994.22
NAGALAND,14070.85,15167.88,16736.27,15389.29
NCT OF DELHI,16673.01,17746.69,18387.79,17709.48
PUNJAB,18539.59,20840.32,23200.49,25050.74
RAJASTHAN,15342.09,16709.98,18265.23,19701.76
SIKKIM,15651.88,16809.80,18101.99,19267.89
TAMIL NADU,21438.90,22914.46,24709.25,24571.01
TRIPURA,18022.14,19140.91,20563.25,19326.66
UTTAR PRADESH,17906.04,18316.04,19808.38,20531.69
UTTARAKHAND,21577.53,23869.98,24772.73,26110.84
WEST BENGAL,18335.66,19258.02,21673.08,21853.27
}\NonCommDALYdata

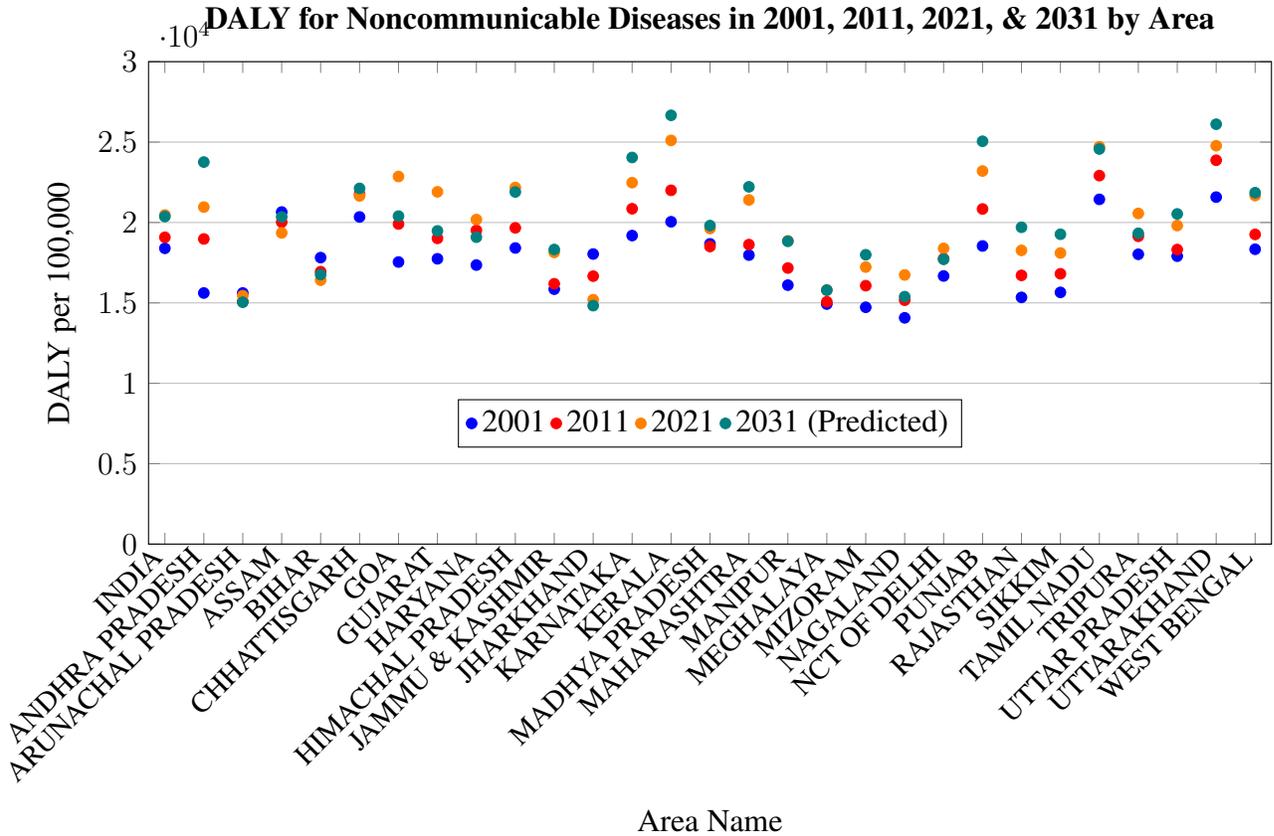
\begin{figure}[h!]
\centering
\begin{tikzpicture}
\begin{axis}[
    title={\textbf{DALY for Noncommunicable Diseases in 2001, 2011, 2021, \& 2031 by Area}},
    xlabel={Area Name},
    ylabel={DALY per 100,000},
    % Define all area names for the symbolic x-axis
    symbolic x coords={
        INDIA, ANDHRA PRADESH, ARUNACHAL PRADESH, ASSAM, BIHAR, CHHATTISGARH, 
        GOA, GUJARAT, HARYANA, HIMACHAL PRADESH, JAMMU \& KASHMIR, JHARKHAND, 
        KARNATAKA, KERALA, MADHYA PRADESH, MAHARASHTRA, MANIPUR, MEGHALAYA, 
        MIZORAM, NAGALAND, NCT OF DELHI, PUNJAB, RAJASTHAN, SIKKIM, TAMIL NADU, 
        TRIPURA, UTTAR PRADESH, UTTARAKHAND, WEST BENGAL
    },
    xtick=data, % Place ticks at each x-coordinate
    xticklabel style={rotate=45, anchor=east, font=\small}, % Rotate labels
    ymin=0,
    ymax=30000, % Set y-axis maximum based on the data
    enlarge x limits=0.015,
    ymajorgrids=true,
    width=\textwidth,
    height=8cm,
    % Configure the legend to be above the plot and horizontal
    legend style={
        at={(0.5, 0.2)},
        anchor=south,
        legend columns=4 
    }
]

% Plot the data for each year with distinct colors
\addplot[only marks, mark=*, mark options={fill=blue, draw=blue}] 
    table [x=Area, y=y2001] {\NonCommDALYdata};
\addlegendentry{2001}

\addplot[only marks, mark=*, mark options={fill=red, draw=red}] 
    table [x=Area, y=y2011] {\NonCommDALYdata};
\addlegendentry{2011}

\addplot[only marks, mark=*, mark options={fill=orange, draw=orange}] 
    table [x=Area, y=y2021] {\NonCommDALYdata};
\addlegendentry{2021}

\addplot[only marks, mark=*, mark options={fill=teal, draw=teal}] 
    table [x=Area, y=y2031] {\NonCommDALYdata};
\addlegendentry{2031 (Predicted)}

\end{axis}
\end{tikzpicture}
\caption{Scatterplot showing DALY per 100,000 for noncommunicable diseases across various areas in India for the years 2001, 2011, 2021, and with predictions for 2031.}
\end{figure}

\subsubsection{Analysis of DALY (Type C - Injury) trends over four decades}
Similar to communicable diseases, Figure 12 also shows a clear and consistent downward trend for the vast majority of states. The visual progression from the higher starting points in 2001 to the lower predicted values in 2031 indicates a sustained, nationwide improvement in reducing the DALYs associated with injuries over the past two decades.

While the trend is consistently negative, the rate of improvement varies by state. Some states, like Gujarat and Jharkhand, show a very steep decline, indicating rapid progress. Others show a more moderate, steady reduction. This variation highlights that the factors influencing injury rates, such as road safety, occupational health, and emergency response, have evolved differently across the country.

This plot also allows for comparison of noncommunicable DALY burdens among states in 2031:\\
Higher-Burden States (DALYs > 3500): Chhattisgarh, Tamil Nadu, Uttarakhand \\
Lower-Burden States (DALYs < 1700): Assam, Gujarat, Jharkhand, NCT of Delhi

% --- Data Table for Injury DALY Values ---
% Data combined from your paper's existing figures and the new PDF
\pgfplotstableread[col sep=comma]{
Area,y2001,y2011,y2021,y2031
INDIA,4854.11,4123.92,3357.39,2671.37
ANDHRA PRADESH,4834.52,3695.19,3045.13,2100.05
ARUNACHAL PRADESH,3496.61,2980.51,2373.71,2078.52
ASSAM,5450.04,4004.60,2782.87,1522.56
BIHAR,4542.15,3712.14,2666.25,1837.19
CHHATTISGARH,5458.89,5674.10,4243.35,3754.38
GOA,3322.94,3082.54,2589.28,2511.22
GUJARAT,6813.52,4078.23,3470.66,1365.81
HARYANA,4470.71,4212.86,3118.47,2729.81
HIMACHAL PRADESH,3617.38,3885.63,3253.94,3343.68
JAMMU \& KASHMIR,4792.69,3340.07,3003.21,1867.19
JHARKHAND,4943.48,3584.96,2246.45,997.78
KARNATAKA,4682.13,4562.52,3609.90,3319.15
KERALA,3776.08,3249.04,2843.31,2439.61
MADHYA PRADESH,5481.87,4830.66,3918.99,3208.18
MAHARASHTRA,4153.84,3435.36,3137.40,2569.21
MANIPUR,3840.64,3611.84,2895.74,3027.25
MEGHALAYA,2589.78,2317.44,1879.03,1760.57
MIZORAM,2888.83,2685.35,2120.11,2055.98
NAGALAND,2986.76,2477.20,2108.92,1885.43
NCT OF DELHI,3325.23,2857.11,2009.04,1604.29
PUNJAB,3879.42,4007.73,3179.04,3224.74
RAJASTHAN,3472.07,3634.52,3076.25,2994.95
SIKKIM,3439.52,3149.65,2388.69,1985.17
TAMIL NADU,6420.64,5748.17,4421.54,3577.36
TRIPURA,4605.24,3797.24,2910.95,2569.56
UTTAR PRADESH,5195.14,4297.29,3810.04,3043.32
UTTARAKHAND,5735.47,5141.56,4656.13,4208.17
WEST BENGAL,4143.29,3673.62,2901.97,2366.25
}\InjuryDALYdata

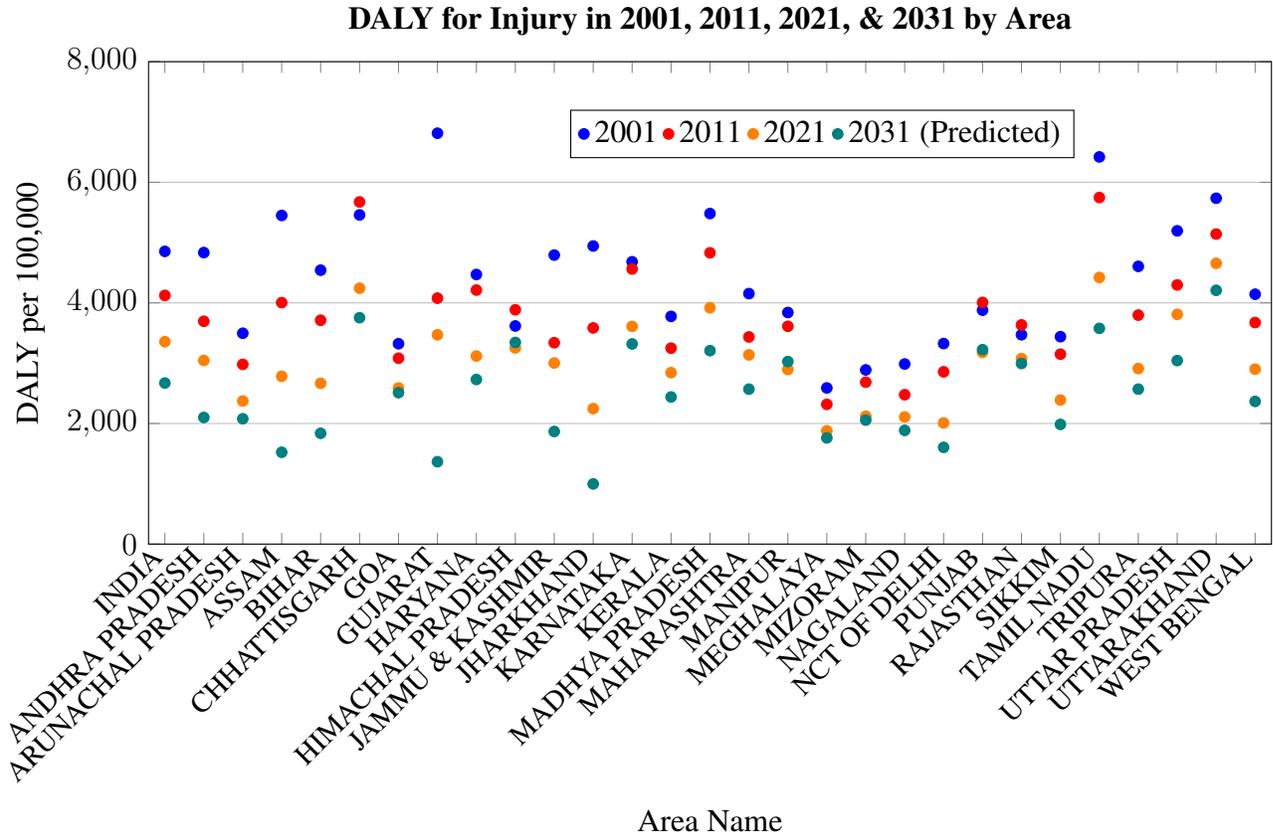
\begin{figure}[h!]
\centering
\begin{tikzpicture}
\begin{axis}[
    title={\textbf{DALY for Injury in 2001, 2011, 2021, \& 2031 by Area}},
    xlabel={Area Name},
    ylabel={DALY per 100,000},
    symbolic x coords={
        INDIA, ANDHRA PRADESH, ARUNACHAL PRADESH, ASSAM, BIHAR, CHHATTISGARH, 
        GOA, GUJARAT, HARYANA, HIMACHAL PRADESH, JAMMU \& KASHMIR, JHARKHAND, 
        KARNATAKA, KERALA, MADHYA PRADESH, MAHARASHTRA, MANIPUR, MEGHALAYA, 
        MIZORAM, NAGALAND, NCT OF DELHI, PUNJAB, RAJASTHAN, SIKKIM, TAMIL NADU, 
        TRIPURA, UTTAR PRADESH, UTTARAKHAND, WEST BENGAL
    },
    xtick=data,
    xticklabel style={rotate=45, anchor=east, font=\small},
    ymin=0,
    ymax=8000, % Set y-axis maximum to fit all data points
    enlarge x limits=0.015,
    ymajorgrids=true,
    width=\textwidth,
    height=8cm,
    legend style={
        at={(0.6, 0.8)},
        anchor=south,
        legend columns=4 
    }
]

% Plot the data for each year with distinct colors
\addplot[only marks, mark=*, mark options={fill=blue, draw=blue}] 
    table [x=Area, y=y2001] {\InjuryDALYdata};
\addlegendentry{2001}

\addplot[only marks, mark=*, mark options={fill=red, draw=red}] 
    table [x=Area, y=y2011] {\InjuryDALYdata};
\addlegendentry{2011}

\addplot[only marks, mark=*, mark options={fill=orange, draw=orange}] 
    table [x=Area, y=y2021] {\InjuryDALYdata};
\addlegendentry{2021}

\addplot[only marks, mark=*, mark options={fill=teal, draw=teal}] 
    table [x=Area, y=y2031] {\InjuryDALYdata};
\addlegendentry{2031 (Predicted)}

\end{axis}
\end{tikzpicture}
\caption{Scatterplot showing DALY per 100,000 for injuries across various areas in India for the years 2001, 2011, 2021, and with predictions for 2031.}
\end{figure}

\section{Key Findings}
The analysis on the changing landscape of disability in India (2001 to 2031), leads to the following key findings .

1. A strong inverse correlation exists between the HDI and the burden of communicable diseases. As states’ HDI values increased, the DALYs for communicable diseases decreased consistently and significantly. The peak DALY dropped from a rate of 38,000 per 100,000 in 2001 to below 18,000 by 2021. These results demonstrate the clear impact that socioeconomic development has on controlling infectious diseases.

2. India is undergoing the epidemiological transition. This public health theory describes a shift in a country’s disease patterns from infectious diseases to chronic, noncommunicable diseases as it develops. The decline in communicable diseases is the first stage of this transition. The second stage is characterized by the rise of noncommunicable diseases which is strongly supported by the data. The DALYs for noncommunicable diseases had no correlation with rising HDIs between 2001 and 2011, and even had a slight positive correlation by 2021. This is a direct result of longer life expectancies and lifestyle changes associated with development.

3. Injury-related DALYs show a moderate but consistent decline over the decades. Injury DALYs also have a weaker correlation with HDI than that of communicable diseases. Over time, the DALY values for injury became more tightly clustered, with the range dropping from 2,500-7,000 to just 2,000-4,700 by 2021. Our analysis suggests that while general development contributes to decreased injury burdens, the relationship is influenced by a more complex set of factors.

4. Gender disparity in disability persists. Males are being overrepresented in the disabled population, with both the 2001 and 2011 censuses having higher ratios of disabled males to females than the overall population ratio. This disparity can be attributed to two reasons. One, the social stigma of disability among women causes underreporting, and two, men have higher occupational risks in physically demanding jobs.

5. 2031 predictions show continued but inconsistent progress. The nation’s overall health, standards of living, and education are predicted to improve, shown by the national HDI rising to 0.709. Injury and communicable disease DALYs are predicted to decline, but noncommunicable DALYs are forecasted to increase. Accounting for these trends, the focus in health policy and legislation must be shifted from overall infectious diseases to the prevention and management of chronic conditions, particularly surrounding managing diseases arising from the modern lifestyle. 

There are a few learnings to be taken from these results. Firstly, noncommunicable diseases will become a larger risk in the coming decades. There must be increased focus on preventing chronic conditions (e.g. diabetes) to prevent against the predicted increasing rates. Secondly, better road safety and occupational health programs must be developed to tackle injury burdens. Next, the decreased communicable disease burdens must be continued through maintained sanitation infrastructure and health systems. And finally, systemic changes must occur to correct existing gender disparities. Men face high occupational injury rates, while disabled women remain invisible due to underreporting.

Lessening these overall disability burdens will require a multipronged strategy. Beyond expanded healthcare investment, engineers must also design infrastructure, design preventive regulations, and social workers are needed to ensure fair access.

\section{Conclusion}
This paper analyzes the evolution of disability burdens in India over two decades (2001-2021). Machine learning was then utilized to extrapolate future trends in DALY, HDI, and gender disparities. The results displayed India’s epidemiological shift and changes in each disability type’s burden. Increases in HDI resulted in decreased communicable disease and injury-related DALYs. But we also see that noncommunicable diseases have remained steady or increased. These changes display the effects of improvements in road safety, legislative policies, and emergency services. 

Gender disparities are still prevalent across India, with men being overrepresented in the disabled population, though at consistently decreasing rates. The 2031 machine learning predictions show these continued trends into 2031. This is likely due to occupational hazards and social stigma leading to underreporting of female disability. This gives the imperative to policy changes. India must prioritize public education, preventative care, and address gender-based differences in healthcare. Changes in gender disability ratios are contingent on improving these factors.

We hope to extend this work in the future. The dataset should include more frequent and recent time points, allowing for nonlinear models that better capture health trends. We also hope to utilize more complete data, including all union territories and Odisha (formerly Orissa).  With these changes, we would be better able to explore how disability affects other social factors.

\section{Scope \& Limitation}
This study analyzed disability burden measured by Disability-Adjusted-Life-Years and Human Development Index over three decades. Though the data in this study was chosen to be detailed and provide a complete understanding of disease burden in India, this study still has several limitations.\\

The authors would like to highlight the following limitations for this study:

Data Granularity: Each data point in this study was collected a decade apart (2001, 2011, 2021), which may not capture yearly fluctuations or the short-term impact of policy changes. Therefore, this study should only be used as a general long-term estimate for future trends.

Geographic Coverage: All Union Territories were excluded due to the unavailability of DALY data and the state of Odisha due to its lack of 2001 census data. This means the findings may not cover the entirety of India.

Reliance on Global Health Estimates: DALY metrics were sourced from the Institute for Health Metrics and Evaluation (IHME), the organization that produces the Global Burden of Disease study. While the metrics the IHME produces have become very important tools for understanding global health, the methodologies for producing global estimates have been noted to sometimes obscure underlying assumptions and knowledge gaps in the data for specific countries or diseases (Tichenor \& Sridhar, 2020). Therefore, due to DALY values being globally standardized health data, the data used in this paper should be understood as estimates that may not fully capture the specific local context.

Prediction Model: The machine learning model used for various HDI and DALY forecasts is a simplified linear extrapolation of past trends. It does not account for potential non-linear trends or the impacts of any unforeseen events.

2021 Census Data: A gender proportion study for 2021 was not undertaken because the 2021 Census of India was postponed and has not yet been conducted. This creates a gap in the most recent analysis of gender disparities.

Reporting Biases: The analysis of gender disparities relies on census data, which may be influenced by the underreporting of disabilities among women due to social stigma (Bhardwaj \& Kumar, 2025), causing skewed gender ratio results.

\section*{Acknowledgment}
The authors would like to thank Dr. Nived Chebrolu, University of Oxford, for his invaluable insights in machine learning predictions. We also acknowledge the Cambridge Centre for International Research for providing the guidance and opportunity to conduct this research.

\section*{Funding}
This research received no specific grant from any funding agency in the public, commercial, or not-for-profit sectors.

\section*{Data Availability}
The dataset used and predicted in this study is uploaded for public access in the GitHub repository. The relevant coding and Machine learning model can also be accessed here. The full dataset can be found at \url{https://github.com/Arunimad/India_DALY_HDI_prediction.git} 

\section*{Conflict of Interest}
The authors declare that they have no financial interests or personal relationships that could have influenced the work reported in this paper.

\nocite{*} % This command tells LaTeX to include EVERY entry from your .bib file
\FloatBarrier % <--- ADD THIS LINE
\printbibliography[title={References}]

\end{document}